\documentclass[prc,aps,twocolumn,float,floatfix,nofootinbib,superscriptaddress,byrevtex,showpacs]{revtex4}
\usepackage{amssymb}
\usepackage{amsmath}
\usepackage{amsfonts}
\usepackage{epsfig}
\usepackage{color}
\newcommand{\etal}{\emph{et al.}}
\newcommand{\nuc}[2]{$^{#1}$\textrm{#2}}

\begin{document}
\title{Collectivity-induced quenching of signatures for shell closures}
\author{M. Bender}
\affiliation{Universit{\'e} Bordeaux,
             Centre d'Etudes Nucl{\'e}aires de Bordeaux Gradignan, UMR5797,
             F-33175 Gradignan, France}
\affiliation{CNRS/IN2P3,
             Centre d'Etudes Nucl{\'e}aires de Bordeaux Gradignan, UMR5797,
             F-33175 Gradignan, France}
\author{G. F. Bertsch}
\affiliation{Department of Physics and Institute for Nuclear Theory,
             Box 351560, University of Washington, Seattle, WA 98195}
\author{P.-H. Heenen}
\affiliation{Service de Physique Nucl\'eaire Th\'eorique,
             Universit\'e Libre de Bruxelles, C.P. 229, B-1050 Bruxelles,
             Belgium}
\affiliation{Physics Division,
             Argonne National Laboratory,
             9700 S. Cass Avenue,
             Argonne, IL 60439,
             U.S.A.}
\date{27 August 2008}
%
%
\begin{abstract}
Mass differences are an often used as signature and measure for shell
closure. Using the angular-momentum projected Generator Coordinate
Method and the Skyrme interaction SLy4, we analyze the modification
of mass differences due to static deformation and dynamic fluctuations
around the mean-field ground state.
\end{abstract}
\pacs{21.10.Dr, 
      21.60.Jz, 
}
\maketitle
%
%
\section{Introduction}

How the magic numbers that are well established along the line of
stability evolve when moving away towards nuclei with large
neutron or proton excess is still an unsettled question. Major
recent experimental progress now allows one to follow magic
numbers over long isotopic or isotonic chains far from stability.
The data for many observables point to the disappearance of well
established magic numbers and the appearance of new ones in nuclei
which sometimes are still close to stability, and far from the
drip lines.

From a theoretical point of view, the situation is still unclear. The
interactions used in shell model calculations have had to be revised in
order to be able to reproduce the new experimental data and to describe
the evolution of shells, especially in nuclei in the
$pf$~shell~\cite{Pov01a,Hon04a,Hon05a,Din05aE,Pri01a,Jan02a}.

An analysis in terms of the self-consistent mean-field method performed
well before most experimental data became available~\cite{Dob94a}
has shown that, when approaching the drip lines, the increasing diffuseness
of the nuclear surface for neutrons and the interaction between bound
orbitals and the continuum affect the shape of the mean-field potentials
of protons and neutrons. As a consequence, shell structure is modified,
known gaps are quenched, and new ones might open.
However, the coupling between bound orbitals and the continuum can be
expected to be an important mechanism for nuclei in the immediate vicinity
of the drip line only, when the Fermi energy is approaching zero. With
the exception of the case of the lightest nuclei with mass below 30,
there are no experimental data for such nuclei, and it is not obvious that
the coupling with the continuum is an effect which has an impact on the
known anomalies in shell structure. By contrast, the diffuseness of
the neutron density might play a role in unstable nuclei, mainly
through a reduction of the spin-orbit potentials for both protons and
neutrons~\cite{Eif95a,Lal98a}.
However, this effect can be expected to appear gradually,
and cannot not be responsible for those anomalies
in shell structure where nuclear structure is suddenly changing when
adding or removing a few nucleons only.

There are several observables that are used as possible signatures
to put the evolution of shell closures in evidence when following
isotopic or isotonic chains. From a theoretical point of view, the
simplest one is the two-nucleon separation energy in even-even
nuclei, which can be easily calculated from the total binding
energy. A jump in the two-nucleon separation energy  is a direct
indication of a sudden increase in the ground-state binding energy
of a given nucleus, and often has a shell closure as its origin.
Other quantities that are frequently analyzed and which are not
directly related to the ground-state properties, but to the gaps
that separate the different shells are provided by excitation
spectra in even and odd-mass nuclei. At least in the context of
mean-field-based models \cite{RMP}, these are more difficult to
calculate and have larger associated theoretical uncertainties.
Still, any conclusion about the evolution of shell structure
requires clear clear trends from a variety of observables that
probe complementary aspects. In particular, it should be supported
also by data on electromagnetic moments and transition
probabilities which provide information on the structure of the
ground state and the excited-state wave functions.

Although there is no formal justification for this practice, the
discussion of shell evolution is often conducted by theorists and
experimentalists alike as if the following three quantities were equivalent:
\begin{enumerate}
\item
    single-particle energies in a spherical self-consistent mean
    field. They are then the eigenvalues of a single-particle Hamiltonian
    and the corresponding wave function is a single Slater determinant,
    i.e.\, a pure state;
\item
    effective single-particle energies as defined within the interacting
    shell model. They should in principle result from an average over a huge number
     of configurations~\cite{Cau02a,Cau05a,Smi06a,Bar70a,Sig07a,Kay08a,Ume08a};
\item
    two-nucleon separation energies.
\end{enumerate}
The last quantity is the only one which can be directly related to
experimental data. Let us recall its definition:
\begin{eqnarray}
\label{eq:s2p}
S_{2p} (Z,N)
& = & E(Z-2,N) - E(Z,N)
      \\
\label{eq:s2n}
S_{2n} (Z,N)
& = & E(Z,N-2) - E(Z,N)  \, ,
\end{eqnarray}
where $E(Z,N)$ is the total binding energy of the nucleus $Z,N$.

In the highly idealized pure HF case where the Koopman theorem is
valid~\cite{Koh74a,Suh06a}, the two-nucleon separation energy is
equal to $-2$ times the energy of the doubly-degenerate orbital
occupied by the two additional nucleons. This is never exactly the
case in nuclear physics. First, pairing effects require a change
in the energy of the Fermi level, resulting through self
consistency to a change of all orbits. Even without pairing
correlations, the two extra particles have an effect on the core
and may induce at least a small rearrangement of all the orbits,
or in the worst case, the appearance of deformations.

It is the purpose of this paper to analyze the role of static and
dynamical quadrupole correlations on the systematics of
$S_{2q}(Z,N)$ in the context of mean-field based methods. Our
analysis relies on the results of Ref.~\cite{Ben05a,Ben06a} where
systematic calculations of the masses of even-even nuclei were
presented. The method allows one to introduce the correlations due
to symmetry restorations (particle number and angular momentum)
performed on axial mean-field  configurations and to add a mixing
of the projected configurations with respect to the axial
quadrupole moment in the framework of the discretized Generator
Coordinate Method (GCM). Since the starting point of the method is
a mean-field calculation, the results give access at the same time
to the spherical single-particle spectra of all even-even nuclei
and to the two-nucleon separation energy of states including
beyond mean-field correlations.

In the next section, we recall the main features of our method. We
then discuss in details the evolution of the two-nucleon
separation energy along $Z$ or $N=50$. Finally, we compare the
results obtained for all isotopic and isotonic chains with the
three different steps of our method to the experimental data.

%
%
\section{The model}

The method used to calculate binding energies for the ground states of
even nuclei is described in detail in Refs.~\cite{Ben05a,Ben06a} (referred
to as paper~I in the following). In this analysis, we use the
energies as tabulated in~\cite{EPAPS}.

To summarize the key features of our method, its starting point is
a set of mean-field calculations including a constraint on the
axial quadrupole moment. As effective interaction we employ a
Skyrme energy density functional, the SLy4
parametrization~\cite{Cha98a}, for the mean-field channel, and a
density-dependent, zero-range interaction in the pairing channel.
Two sets of correlations beyond the mean-field are introduced.
First, the deformed wave functions are projected on both fixed
particle numbers and on angular momentum. This projection
corresponds to mixing degenerate mean-field wave functions which
differ by a spatial rotation and to introduce fluctuations around
the orientation of the mean-field. In a collective model
terminology, these correlations would be called rotational
correlations. A second step of our method consists in the mixing
of projected mean-field wave functions as a function of their
axial quadrupole moment. This corresponds to a vibrational
correction in the collective model language. Our final wave
function has the form:
\begin{equation}
| J M \nu \rangle
= \sum_{q} f_{J,\nu} \hat{P}^J_{M0} \hat{P}_N \hat{P}_Z | q \rangle
\, .
\end{equation}
The ket $| q \rangle$ is a (paired) mean-field state of  axial
quadrupole deformation $q$. The particle numbers and angular
momentum quantum number are restored thanks to the operators
$\hat{P}_N$ and $\hat{P}_Z$ which project on good neutron and
proton numbers, and $\hat{P}^J_{M0}$ which projects on
angular-momentum $J$ with the $z$ component $M$ in the laboratory
frame \cite{Rin80a}. The projected wave functions are labeled by
the index $q$ which indicates from which mean-field state they are
obtained, although an intrinsic quadrupole moment has no direct
physical meaning for a wave function with good angular momentum
and can only be defined with the help of the collective model, in
the same way as it is from experimental data. The weights
$f_{J,\nu} (q)$ defining the mixing of the projected wave
functions with respect to $q$ are obtained by variation of the
total energy.

The mean-field wave functions are generated by treating the
pairing correlations using the Lipkin-Nogami (LN) prescription.
This enforces the presence of pairing in all nuclei and for all
quadrupole moments, even at magic numbers, which is necessary to
ensure the continuity of the mean-field wave functions required to
perform the configuration mixing. Binding energies, however, are
always recalculated after a projection on particle
number~\cite{Ben06a}.

We stress that there are no assumptions made in the model about the
amplitude of
the quadrupole fluctuations introduced into the calculations. Depending on the
structure of a nucleus, this amplitude either corresponds to a small vibration
around a pronounced minimum, to a large-amplitude motion in a soft and wide
potential well, or to the mixing of several states around coexisting minima in
the deformation energy surface. Let us recall also that our method is fully
variational and that the energy of the ground state is lowered at
each successive  stage of the calculation.

Finally, one must note that the Skyrme interaction is not a force,
but an energy
density functional. It requires some care to be used beyond a mean-field approach,
and there are still some open questions concerning this long-standing practice.
The density-dependent terms are generalized using the standard prescription that
the density is replaced by the transition density in the density-dependent terms.
This is the only prescription that guarantees various consistency requirements
of the energy functional~\cite{Rod02a,Rob07a}. However, this procedure may lead
to problems that have been put into evidence recently~\cite{Ang01a,Dob07a,Lac08a}.
The numerical procedure that we have used (see Paper~I for details) appears
to be safe in this respect and no dramatic signs of problems have shown up when
varying the number of discretization  points in deformation. In any case, since
we are interested here in trends as a function of $N$ and $Z$, errors
of the order of 200 to 300~keV would not affect our conclusions.

We will now compare results obtained from three wave functions
that successively add quadrupole correlations:
\begin{enumerate}
\item
spherical mean field states $| q=0 \rangle$;
\item
the mean-field minimum in the space of axial reflection-symmetric
deformations  $| q_{\text{min}} \rangle$, which might be spherical;
\item
the ground state obtained after configuration mixing
by the generator coordinate method of $J=0$ projected axial quadrupole.
We refer to these wave functions in the following as projected GCM.
\end{enumerate}
%

%
%
\section{Results}

We will first illustrate the key points of the analysis of
two-nucleon separation energies as measures
of shell closures for the proton $Z=50$ and neutron $N=50$
shells. In a second step, we will then analyze the global
systematics of two-nucleon separation energies for even-even
nuclei across the nuclear chart.

Recent experimental progress has provided many new direct,
high-precision measurements of nuclear
masses~\cite{newmasses,Hak08aM}, which allow comparison of
calculations with data along chains that extend to exotic nuclei.
Experimental masses used here are either taken from
AME2003~\cite{Aud03aM}, where we exclude extrapolated values, or
from Ref.~\cite{newmasses,Hak08aM} when error bars are smaller or
when the AME2003 values from indirect measurements do not agree
with new directly measured ones.

%
%
\subsection{The $Z=50$ isotopic chain}

\begin{figure}[t!]
\includegraphics{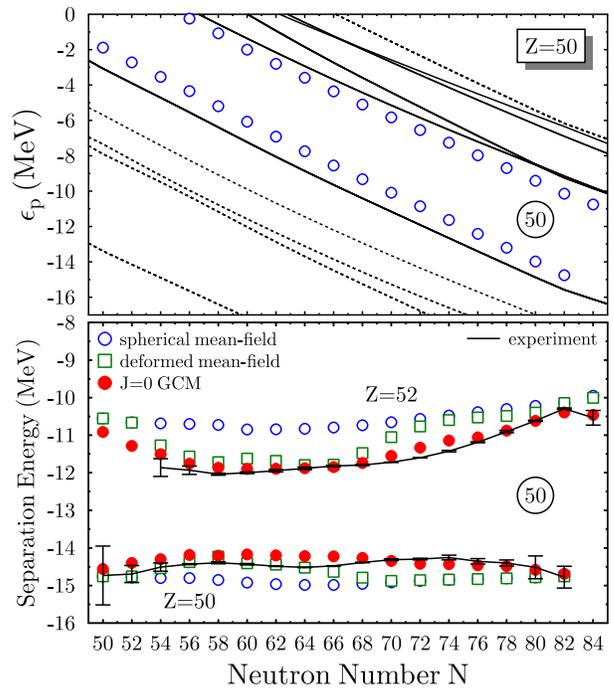}
\caption{\label{fig:s2p:z50}
(Color online) Top: Single-particle spectra of the protons for the
chain of Sn $(Z=50)$ isotopes and $-1/2$ times the two-proton
separation energy for the $Z=50$ and $Z=52$ isotopic chains (see
text). Bottom: Comparison of the gaps between $-S_{2p}(Z=50,N)/2$
and $-S_{2p}(Z=52,N)/2$ obtained at three levels of approximation
with the experimental data. A linear trend equal to
$(N-82)\,[S_{2p}(Z=50,N=50)-S_{2p}(Z=50,N=82)]/2$ using the spherical
mean-field $S_{2p}$ values has been subtracted from all $-S_{2p}(Z,N)/2$,
such that the distance between the curves for a given $N$ is unchanged.
In particular, the plotted values at $N=82$ are not affected, and the
plotted spherical mean-field values for $(Z=50,N=50)$ and $(Z=50,N=82)$
are identical.
}
\end{figure}

Figure \ref{fig:s2p:z50} illustrates the difficulties encountered
when relating two-nucleon separation energies to the gaps in
the single-particle spectrum with the example of the chain of
even-even Sn, $Z=50$, and Te, $Z=52$, isotopes.
The lower panel shows the evolution of the
eigenvalues of the spherical mean-field single-particle Hamiltonian
$\epsilon_p$. Note that all Sn isotopes have a spherical ground
state at the mean-field level.

The open blue circles, plotted in both panels, represent the
two-proton separation energies $S_{2p}$ for $N=50$ and $N=52$
from spherical mean-field calculations.

The magnitude of the $Z=50$ gap in the single-particle spectrum is
fairly independent of $N$ in this case. The spacing between the
$-S_{2p}/2$ for $Z=50$ and 52 is a poor measure of the true shell
gap, even when $-S_{2p}/2$ values and the single-particle levels
are calculated in the same framework and the same shape is
enforced  for all nuclei. Indeed, the $-S_{2p}/2$ spacing turns
out to be smaller for all $N$ values. This result can be partly
attributed to our treatment of pairing by the LN prescription.
This method has the effect of moving the single-particle states
away from the Fermi level. However, the amplitude of this change
is of the order of 200 to 300~keV at most and does not account for
the difference between the single particle levels and $-S_{2p}/2$.

Still, the evolution with $N$ of the spherical mean-field $-S_{2p}/2$ values
follows closely that of the single-particle levels:
the difference between their values for $Z=50$ and $Z=52$ is nearly
constant.

In the lower panel of this Fig.~\ref{fig:s2p:z50}, the $-S_{2p}/2$
values for both isotopic chains are shown for spherical, deformed,
and projected GCM calculations together with the experimental
data. A global descending linear trend with $N$ has been taken
out. The $-S_{2p}/2$ values from spherical mean-field calculations
differ the most from the actual data. They are smaller for $Z=50$
and larger for $Z=52$ than the data. In particular, the distance
between the experimental $S_{2p}$ values at $Z=50$ and at $Z=52$
is not constant; it decreases when going away from $N=82$, and
starts to grow when approaching $N=50$.

\begin{figure}[t!]
\includegraphics{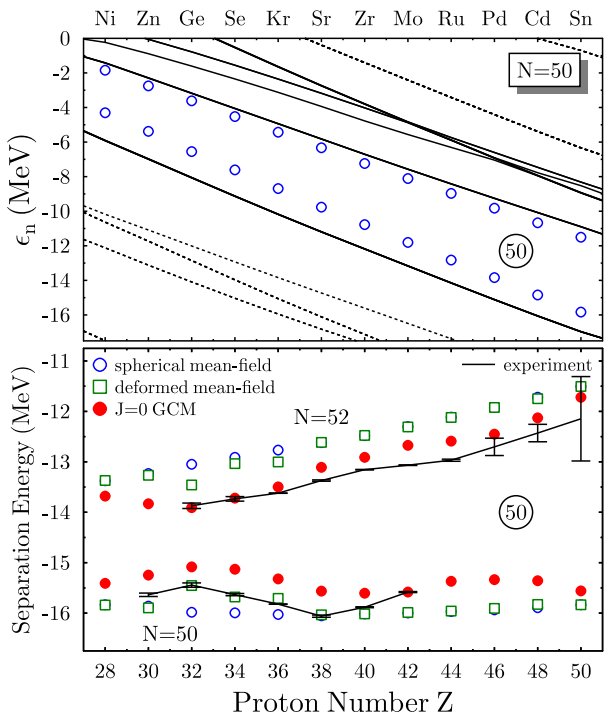}
\caption{\label{fig:s2n:n50}
(Color online)
Same as Fig.~\ref{fig:s2p:z50} but for neutrons. In
the lower panel, a global descending trend $(N-50) \,
[S_{2n}(Z=28,N=50)-S_{2n}(Z=50,N=50)]/2$ using the spherical
mean-field $S_{2n}$ values has been subtracted  from
all $-S_{2n}(Z,N)/2$. The overall reduction of the distance
between $S_{2n}(Z,N=50)$ and $S_{2n}(Z,N=52)$ when approaching
$Z=28$ is clearly visible.
}
\end{figure}

A significant improvement is achieved by allowing the mean-field
to deform. As outlined in paper~I, some of the Cd ($Z=48$) and Te
($Z=52$) isotopes gain up to 2~MeV in binding energy thanks to
deformation, and this increases $S_{2p}$ for \mbox{$Z=50$} and
decreases it for \mbox{$Z=52$}. Despite this effect of
deformation, none of the nuclei involved in the calculation of the
$S_{2p}$ can be classified as well-deformed: they exhibit a
transitional pattern, with soft deformation energy surfaces, and
present shallow deformed minima in some cases. This soft
topography has a significant effect when the fluctuations around
the mean-field minima are taken into account. The Cd and Te
isotopes are usually softer than the Sn ones for a given
$N$-value, in such a way that Cd and Te gain more dynamical
correlation energy. This increases the $S_{2p}$ values for $Z=50$,
but reduces them for \mbox{$Z=52$}.

Specifically, for these two isotopic chains, the change in the
$S_{2p}$ energies brought about by the dynamical quadrupole
correlation energy is often large when the change due to  the
static deformation energy is small and vice versa. The resulting
effect is that the projected GCM curves are much smoother than the
deformed mean-field ones, and closely follow the experimental
data. The only discontinuity left in Fig.~\ref{fig:s2p:z50} is at
the neutron shell closure $N=82$, where the distance between
\mbox{$S_{2p}(N,Z=50)$} and \mbox{$S_{2p}(N,Z=52)$} is largest, a
phenomenon coined as "mutually enhanced magicity" in the
literature~\cite{Schm79a,Zel83a,Lun03a}. From this result one can
conclude that, within the present theoretical framework, the
variation of the gap in the $S_{2p}$ values as a function of $N$
is not related to a change in the underlying spherical shell
structure, but to the variation of the energy contribution brought
by static and dynamic quadrupole correlations to the masses of the
nuclei that enter the calculation of $S_{2p}$. The same
qualitative behavior has been found for proton shells at $Z=82$ in
Refs.~\cite{Ben02a,Ben05a,Ben06a}.

%
%
\subsection{The $N=50$ isotopic chain}

The same analysis for the chains of $N=50$ and 52 isotones is
performed in Figure~\ref{fig:s2n:n50}. Recent mass measurements
\cite{Hak08aM} allow one to follow $S_{2n}$ down to $Z=30$.
Qualitatively, the results are quite similar to those discussed
for the $Z=50$ and 52 chains above, with two notable exceptions.
First, the size of the $N=50$ gap in the single-particle spectrum
is slowly and continuously decreasing from 6.1~MeV in
\nuc{100}{Sn} to 4.4~MeV in \nuc{78}{Ni}, which constitutes "real"
shell quenching and is a consequence of the increasing effect of
the surface on neutron levels that are pushed up in the potential
well. The spin-orbit splitting between the neutron $g_{9/2^+}$ and
$g_{7/2^+}$ levels is slightly increased in our calculation when
going from \nuc{100}{Sn} down to \nuc{84}{Ge}, the lightest
isotone where both are bound. The striking difference with respect
to the protons in the Sn chain discussed above can be related to
the impact of the Coulomb barrier: its presence for protons
suppresses the diffuseness of the wave functions. Second, several
fluctuations are superimposed on this smooth global trend, one of
which slightly opens the gap at the proton sub-shell closures
$Z=38$ and $Z=40$ and closes it for smaller $Z$ values, such that
the calculated gap is smallest for $Z=32$, in agreement with
experiment~\cite{Hak08aM}, when taking also the experimental
$S_{2n}$ value of the odd nucleus ($Z=31$, $N=52$) into account.

A comparison of predictions of various mean-field and other mass
models for the $N=50$ isotonic chain with the most recent data can
be found in Ref.~\cite{Hak08aM}.

%
%
\subsection{Global systematics}

\begin{figure}
\includegraphics[width=8.6cm]{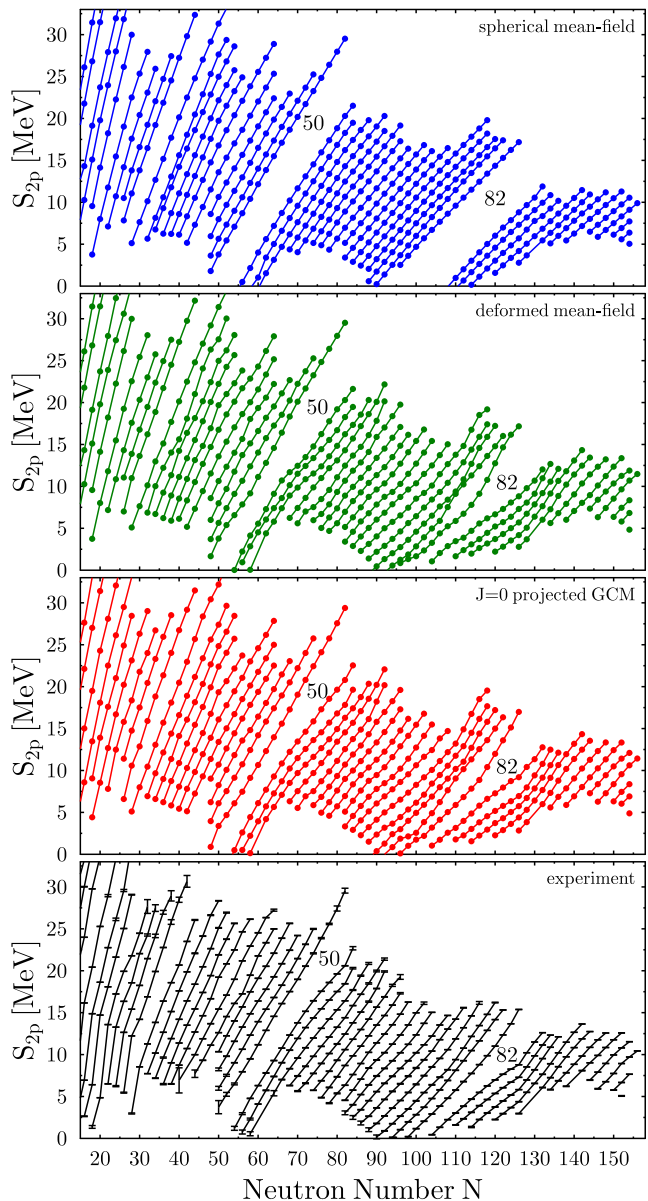}
\caption{\label{fig:s2p:global}
(Color online) Two proton-separation energy $S_{2p}$ for even-even
nuclei. Lines connect nuclei in isotonic chains. }
\end{figure}

\begin{figure}
\includegraphics[width=8.6cm]{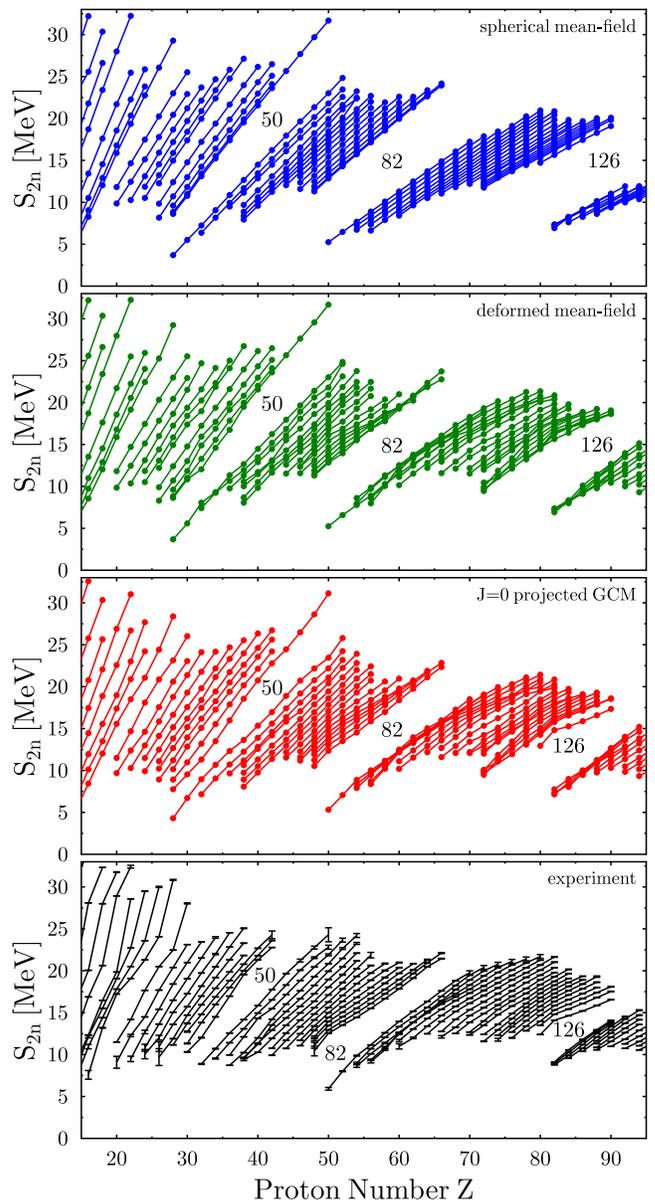}
\caption{\label{fig:s2n:global}
(Color online) Two-neutron separation energy $S_{2n}$ for
even-even nuclei. Lines connect nuclei in isotopic chains }
\end{figure}

Figure~\ref{fig:s2p:global} provides the two-proton separation energy
$S_{2p}$ for isotonic chains (now without the factor $-1/2$)
in all but the lightest even-even nuclei calculated in
\cite{Ben05a,Ben06a}. Together with the experimental data, the
plots give the results for spherical and
deformed mean-field calculations and for the projected GCM ground
states.

The $S_{2p}$ energies from spherical mean-field calculations (top panel)
nearly follow straight lines, with a slope that slowly decreases with mass.
Three gaps, corresponding to the magic numbers $Z=28$, 50 and 82, are
clearly visible. Their widths do not vary significantly with neutron number.
Most importantly, there is no sign of shell quenching in the $S_{2p}$ trends,
consistent with the gaps in the single-particle spectra that remain fairly
constant as well.

The inclusion of static mean-field deformations has a significant effect
on the curves. Compared to the spherical results, the magic gaps in
the $S_{2p}$ values are reduced in many instances. As in the examples
of the $Z=50$ and $N=50$ gaps, the magic nuclei remain spherical, but
not all of their neighbors with $\pm 2$ protons are. The slope of the
$S_{2p}$ curves is also modified between the spherical shells: the energy
gain due to deformation increases $S_{2p}$ above magic numbers and decreases
it below. Moreover, several small local gaps open between the spherical
shell closures, and indicate deformed shell closures.

Correlations beyond mean-field amplify the effects brought by static
deformation: the gaps at the magic numbers are reduced further. Confirming
the discussion for the $Z=50$ chain, deformation and dynamical correlations
reduce the gap in the $S_{2p}$ energies at all shell closures to about
half their value in spherical mean-field calculations. Also, the effect
on energy of dynamical correlations is the largest at shell closures,
where it rapidly changes from nucleus to nucleus, while it is nearly
constant for open shell nuclei~\cite{Ben06a}. It is quite gratifying
to see that adding the correlations brings the calculated values
close to experiment. Note that a particularly strong quenching is
visible for $Z=82$. The three calculations presented here are based on
the same effective interaction which predicts a spherical
single-particle spectrum with gaps much larger than those in the
experimental $S_{2p}$ energies that vary only by a few 100~keV with
the neutron number. Hence, the quenching observed in the experimental
$S_{2p}$ values results from the effects of deformation and of beyond
mean-field correlations on the total binding energy and could
be called "collectivity enhanced" quenching of the two-nucleon separation
energies at shell closures, rather than quenching of the gaps in the spherical
single-particle spectra.

Figure~\ref{fig:s2n:global} shows the two-neutron separation
energies $S_{2n}$ for isotopic chains. Qualitatively, one
finds a similar behavior as for the $S_{2p}$ curves, and the quenching
of the  $S_{2n}$ values from quadrupole correlations is even more pronounced.
However, the overall agreement with experiment of the $S_{2p}$ curves
from the $J=0$ projected GCM is better than the one for $S_{2n}$, as was
already pointed out and discussed in detail in~\cite{Ben05a,Ben06a}.
The Skyrme interaction SLy4 used here (as
all others we tested) seems to systematically overestimate neutron
shell gaps, while proton shell gaps are described better.

The inclusion of beyond-mean field correlations has a smoothing
effect on the variation of the two-particle separation energies as
a function of $N$ or $Z$. The often unrealistic local fluctuations
obtained without these correlations in our calculations or in
those of deformed mean-field calculations by other
groups~\cite{Dob04a}, or from mean-field-based mass models such
as the microscopic-macroscopic method~\cite{Mol95a} or the
Brussels HFB mass fits~\cite{Gor01a}
are to a large extent suppressed. Thus, the correlation energy
added by fluctuations of the wave functions around the mean-field
minima compensates the too abrupt shape changes (spherical to
deformed, or prolate to oblate) obtained within a pure mean-field
approach.

%
%
\section{Summary}

The structural changes that occur in nuclei when going along
isotopic or isotonic chains have been discussed on the basis of
two-nucleon separation energies. We have shown that the effect
that has been referred to as shell quenching and has sometimes
been attributed to a reduction of the spherical gaps far from
stability is adequately described by the introduction of dynamic
collective quadrupole correlations. Much smoother trends with
fewer kinks and discontinuities are obtained than in deformed
mean-field calculations. In many instances, our calculated values
are close to the data. Looking at the shell closures where the
$S_{2q}$ values exhibit discontinuities, static mean-field
deformations and dynamical correlations decrease systematically
the amplitude of these gaps, and reduce them far from stability.
Both effects are not related to a reduction of the spherical shell
structure, rather both underline the importance of fluctuations
around single mean-field configurations for a high-precision
description of nuclear masses. Studies performed in restricted
mass regions~\cite{Ben02a,Ben05a,Ben06a,Jun06a,Fle04a} have
arrived at similar conclusions.

This study addresses also two present-day questions concerning
mean-field models and effective interactions (or density
functionals). For a long time, single particle-energies have been
viewed as being not directly connected with experimental data
because they are strongly renormalized by the coupling to
different kinds of vibrations~\cite{Ham74a,Koh74a,Mah85a,Lit06a}.
However, it has been proposed recently~\cite{Zal08a} that this
statement is too strong and that single-particle energies should
be used in the fitting procedure of energy density functionals.
This point has been convincingly demonstrated using specific data.
Our analysis shows that one must carefully select the data
relevant for adjusting single-particle levels and that the
quenching of 2-particle separation energies should not be related
to single particle properties. Our study also raises the question
of whether all correlations should be included in an
energy-density functional, or if some of them should be treated
explicitly, beyond a mean-field approach. At the minimum, our
study shows the interest of treating explicitly the correlations
associated with fluctuations in collective degrees of freedom.

Finally, let us note that the relation between single-particle
energies and the evolution of 2-particle separation energies along
$N$ or $Z$ is not a simpler problem in the shell model context.
The definition which corresponds the most closely to a spherical
mean-field single particle energy is the effective single-particle
energy of Caurier et al.~\cite{Cau02a,Cau05a}, which includes the
monopole shift of single-particle energies, i.e. the terms of the
interaction that correspond to the spherical mean field. A
definition more directly related to nucleon transfer reactions is
based on centroids of the strength
distribution~\cite{Bar70a,Sig07a,Kay08a,Ume08a}. Both are obtained
from the averaging over correlated states which compromises their
interpretation in terms of a pure configuration, and are neither a
directly observable quantity. The main message of this
contribution is probably that one should not try to carry out too
indirect comparisons with experimental data and that the notion of
single-particle states should be limited to the very few cases
where correlation effects are weak.

%
%
\section*{Acknowledgments}
MB thanks the organizers and participants of the workshop
"Mass Olympics", held at ECT$^\ast$ Trento 26-30 May 2008
for many inspiring presentations and discussions.
This research was supported in parts by the PAI-P5-07 of the
Belgian Office for Scientific Policy, by the U.S.\ Department
of Energy under Grant DE-FG02-00ER41132 (Institute for Nuclear
Theory) and DE-AC02-06CH11357 (ANL).
%
%

\end{document}